\documentclass[aps, reprint, superscriptaddress, showpacs, showkeys]{revtex4-1}

\usepackage{color}
\usepackage{amssymb, bm, amsmath}
\usepackage{graphicx}
\usepackage{epstopdf}
\usepackage{float}
 
\newcommand{\ds}{\displaystyle}
\newcommand{\un}{~\mathrm}
\newcommand{\ddroit}{~\mathrm{d}}
\newcommand{\tepsilon}{\boldsymbol{\epsilon}}
\newcommand{\te}{\boldsymbol{e}}
\newcommand{\tsigma}{\boldsymbol{\sigma}}
\newcommand{\tSigma}{\boldsymbol{\Sigma}}
\newcommand{\tE}{\boldsymbol{E}}
\newcommand{\A}{\mathbb{A}}
\newcommand{\C}{{\mathbb{C}}}
\newcommand{\Cloc}{\mathbb{C}^{\rm loc}}

\newcommand{\Cs}{{\C_{s}}}
\newcommand{\Cp}{{\C_{p}}}

\newcommand{\J}{{\mathbb{J}}}
\newcommand{\K}{{\mathbb{K}}}
\newcommand{\I}{{\mathbb{I}}}
\newcommand{\Id}{{\bf I}}

\newcommand{\vu}{{\boldsymbol{u}}}
\newcommand{\vx}{{\boldsymbol{x}}}
\newcommand{\RH}{{$RH$}}
\newcommand{\MV}[1]{\langle{#1} \rangle}

\definecolor{purple}{rgb}{0.5,0.0,0.5}

\usepackage{lineno}

\begin{document}

\title{Highly porous  layers of silica nano-spheres sintered by  drying: 
 Scaling up of the elastic  properties from the beads to the macroscopic mechanical properties}

\author{A. Lesaine}
\affiliation{Laboratoire FAST, Univ. Paris-Sud, CNRS, Universit\'e Paris-Saclay, F-91405, Orsay, France.}
\affiliation{SPEC, CEA, CNRS, Universit\'e Paris-Saclay, 91191 Gif-sur-Yvette, France.}
\author{D. Bonamy}
\affiliation{SPEC, CEA, CNRS, Universit\'e Paris-Saclay, 91191 Gif-sur-Yvette, France.}
\author{G. Gauthier}
\affiliation{Laboratoire FAST, Univ. Paris-Sud, CNRS, Universit\'e Paris-Saclay, F-91405, Orsay, France.}
\author{C. L. Rountree}
\affiliation{SPEC, CEA, CNRS, Universit\'e Paris-Saclay, 91191 Gif-sur-Yvette, France.}
\author{V. Lazarus}
\affiliation{Laboratoire FAST, Univ. Paris-Sud, CNRS, Universit\'e Paris-Saclay, F-91405, Orsay, France.}
\affiliation{IMSIA, ENSTA ParisTech, CNRS, CEA, EDF, Universit\'e Paris-Saclay, 828 bd des Mar\'echaux, 91762 Palaiseau Cedex, France.}

%

\begin{abstract}

Layers obtained by drying a colloidal dispersion of silica spheres are found to be
a good benchmark 
to test the elastic behaviour of porous media, in the challenging case of high porosities and nano-sized microstructures.
Classically used for these systems, 
Kendall's approach explicitely considers the effect of surface adhesive forces onto the contact area between the particles. 
This approach provides the Young's modulus using a single adjustable parameter (the adhesion energy) but provides no further information on the tensorial nature and possible anisotropy of elasticity. 
On the other hand, homogenization approaches ({\it e.g.} rule of mixtures, Eshelby, Mori-Tanaka and self-consistent schemes), based on continuum mechanics and asymptotic analysis, provide the stiffness tensor from the knowledge of the porosity and the elastic constants of the beads.
Herein, the self-consistent scheme accurately predicts both bulk and shear moduli, with no adjustable parameter, provided the porosity is less than 35$\%$, for layers composed of particles as small as 15 nm in diameter.
Conversely, Kendall's approach is found to predict the Young's modulus over the full porosity range. Moreover, 
  the adhesion energy in Kendall's model has to be adjusted to a value of the order of the fracture energy of the particle material. This {suggests}
that sintering during drying leads to the formation of covalent siloxane bonds between the particles.
\end{abstract}



\maketitle

\section{Introduction}
Porous materials formed by {cohesive beads} are commonly found in nature (sandstones, sedimentary rocks, opals, soils...) and in industry (ceramics, pharmaceutical pills, filter cakes, {photonic materials}, paintings...). Sintering between beads, {whether resulting from evaporation, heat or compression}, confers an overall
cohesion and solid behavior to the material. 
For engineering purposes, it is of utmost importance to relate the mechanical properties at the macroscale to 
microscale behavior, whatever the constitutive relations of the components (elastic, plastic, viscoplastic...).  This constitutes a broad field  \cite{SanZao87Wil, Pon02_Cours_Homogeneisation, DorKonUlm06}; this paper only addresses the linear elastic part.

When looking for the equivalent elasticity of a packing of cohesive grains, a first possibility is {to exploit the analogy between scalar elasticity and scalar electricity \cite{JerCosChe82} and sketch the material as a network of resistances. Effective medium theory \cite{Brug35,Lan52,Kir71} then permits, from the particle coordination number and the density probability function of contact resistance, to compute the  effective resistance and subsequently the elastic modulus of the packed system}.  However, these two  parameters  are difficult to assess, and this approach fails to take into account the inherent tensorial nature of elasticity.
Even in the simple case of an isotropic solid, two parameters ({\it e.g.} bulk and shear moduli) are necessary to fully describe the elastic behavior of the material.
Therefore, it is more appropriate to use  homogenization methods  rigorously derived within the framework of continuum mechanics and multi-scale asymptotic analysis \cite{SanZao87Wil, Pon02_Cours_Homogeneisation, DorKonUlm06}. 
Finally, when the particle size becomes submicrometric, adhesive surface forces are expected to become relevant. 
{Kendall's approach explicitely takes these forces into account. Herein, it proves to be a relevant framework to cast the problem into.} To the best of our knowledge, only a few papers report quantitative  comparisons between these theoretical approaches  and experimental ones (see {\it e.g.} \cite{Ken88, BonChaRou05, AshHab90} for past attempts).

{The study herein proposes nanoporous materials {obtained} by drying a monodisperse aqueous colloidal suspension consisting of nanometer-sized silica spheres (Ludox HS-40) as a benchmark medium to test models against.} 
During drying, water evaporation brings the particles into contact and transforms the initially liquid dispersion into a  solid {layer} constituted of self-organized sintered particles
(Sec.~\ref{se:exp}).  
{Controlling the drying rate provides a simple way to {modulate} the porosity of the dried material \cite{PirLazGau16}. 
Both bulk and shear elastic constants are measured by ultrasound methods.}
Since the goal here is to bridge the continuum mechanics and soft matter communities \cite{GoeNakDut15DessicationCracks, BirYunRee17, SibPau17}, we review the basic ingredients involved in
classical homogenization schemes: the rule of mixtures and the Eshelby, Mori-Tanaka and self-consistent schemes (Sec.~\ref{se:theory}), and in Kendall's approach (Sec.~\ref{se:kendall}).  Sec.~\ref{se:comp} compares the theoretical predictions with the experimental data {and Sec.~\ref{se:discussion} discusses the results}. 

The self-consistent approach accurately predicts both the shear and bulk moduli with no ajustable parameters, as long as the porosity is sufficiently small (less than $35\%$). 
Conversely, Kendall's approach predicts the variations of the Young's modulus with porosity over the full range, provided that the adhesion energy is properly adjusted. The fitted value is found to be surprisingly high when compared to the values usually considered in this kind of problem \citep{CleGoeRou13, BirYunRee17}: It falls very close to the fracture energy, {\it i.e.}~the energy required to break the covalent siloxane bonds, showing that {drying colloidal suspensions enables the formation of strong covalent siloxane bonds between particles.} This apparent discrepancy with the literature is discussed.

\section{Experimental methods}
\label{se:exp}

\subsection{Sample preparation}
\begin{figure}[h!]
\centering
\includegraphics[width=0.7\columnwidth]{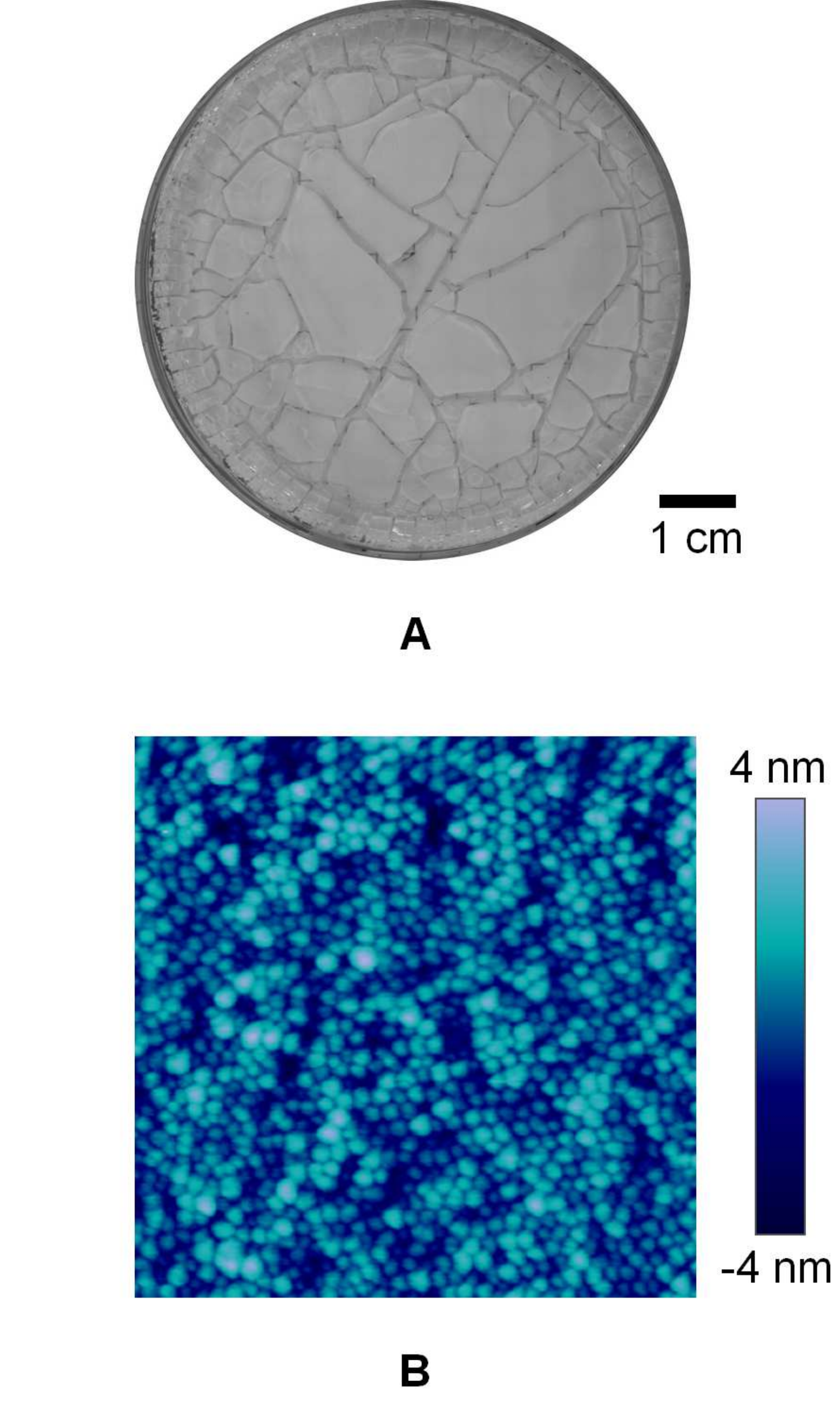}
\centering
\caption{(A) Crack patterns of a dried colloidal layer. (B) Topographical AFM image of the top surface of the layer (scan size $500 \times 500\un{nm}^2$). The sample pictured here was dried at {\RH} = $50\%$.}
\label{fig:RH50_layer}
\end{figure}

The study herein is a follow-up of a previous study \cite{PirLazGau16}. It uses Ludox HS-40, an aqueous dispersion (initial mass concentration $\phi_m \sim 40 \%$) of  silica spheres, commonly used as a model colloidal suspension \cite{DiGDavMit12, BouGioPau15, CabLiArt16, BirYunRee17, SibPau17}. According to SAXS measurements of the form factor, silica beads have an average diameter $a = 16.2 \un{nm}$ with a relative polydispersity $\sigma = 14\%$. In all the study the same batch (number STBF8427V) of Ludox was used.
The porous material is obtained from $25~\mathrm{g}$ of suspension poured into a Petri dish {($3.5 \un{cm}$ radius), which corresponds to an initial liquid height of $5\un{mm}$}.
The system is then left to dry at room temperature ($T=25\pm2\,^\circ\mathrm{C}$) in an enclosure with  constant relative humidity {(\RH)} until the evaporation {ceases}. {This process typically lasts between 2 and 14 days depending on the {{\RH} value}.} 
As water evaporates, the beads {come in contact,} sinter {and form} a  porous solid material of final thickness $h \sim 2 \un{mm}$.

Some desiccation cracks \cite{GoeNakDut15DessicationCracks} appear during this drying process (Fig.~\ref{fig:RH50_layer}A). 
These fractures are due to the {shrinkage}  of the layer induced by evaporation and {impeded} by the substrate ({\it i.e.}~bottom of the Petri dish) \cite{CheLaz13}. These cracks  divide the layer into smaller morsels \cite{LazPau11desc, Laz17}. {The size of the morsels increases with higher {\RH}, that is slower drying.} 

\subsection{Properties at the bead scale}
Due to the nanometric size of the beads, their density ($\rho_{s}$) and bulk ($k_{s}$) and shear ($\mu_{s}$) elastic moduli (or equivalently Young's modulus ($E_{s}$) and Poisson's ratio ($\nu_{s}$)) cannot be directly measured.
Thus, the bead properties are assumed to be equivalent to the bulk properties of pure silica. Specifically, the spec values of pure silica (Corning 7980 standard grade) are used for these parameters \cite{DC7980}:
$\rho_{s}=2.20 \un{g/cm}^{3}$, $k_{s}=35.4 \un{GPa}$, $\mu_{s}=31.4 \un{GPa}$, $E_s=72.7 \un{GPa}$, $\nu_s=0.16$. 

In order to {scale up} the properties of the beads at the macroscale, some knowledge of the particle arrangement at the microscale is necessary. 
Imaging the top of the morsels via an Atomic Force Microscope (AFM) provides the structure of the particle packing. Fig.~\ref{fig:RH50_layer}B gives an example of the particle arrangement for $RH = 50\%$. This image shows that the arrangement is  {on average} homogeneous and isotropic along the surface, at  larger scales. Henceforth, it is assumed that it remains true along the third direction. 
The agreement between the experiments and the models (Sec.~\ref{se:comp}) will  validate this hypothesis, {\it a posteriori}.

 \subsection{Sample properties at the macroscopic scale}

\begin{figure}[h!]
\centering
\includegraphics[width=0.9\columnwidth]{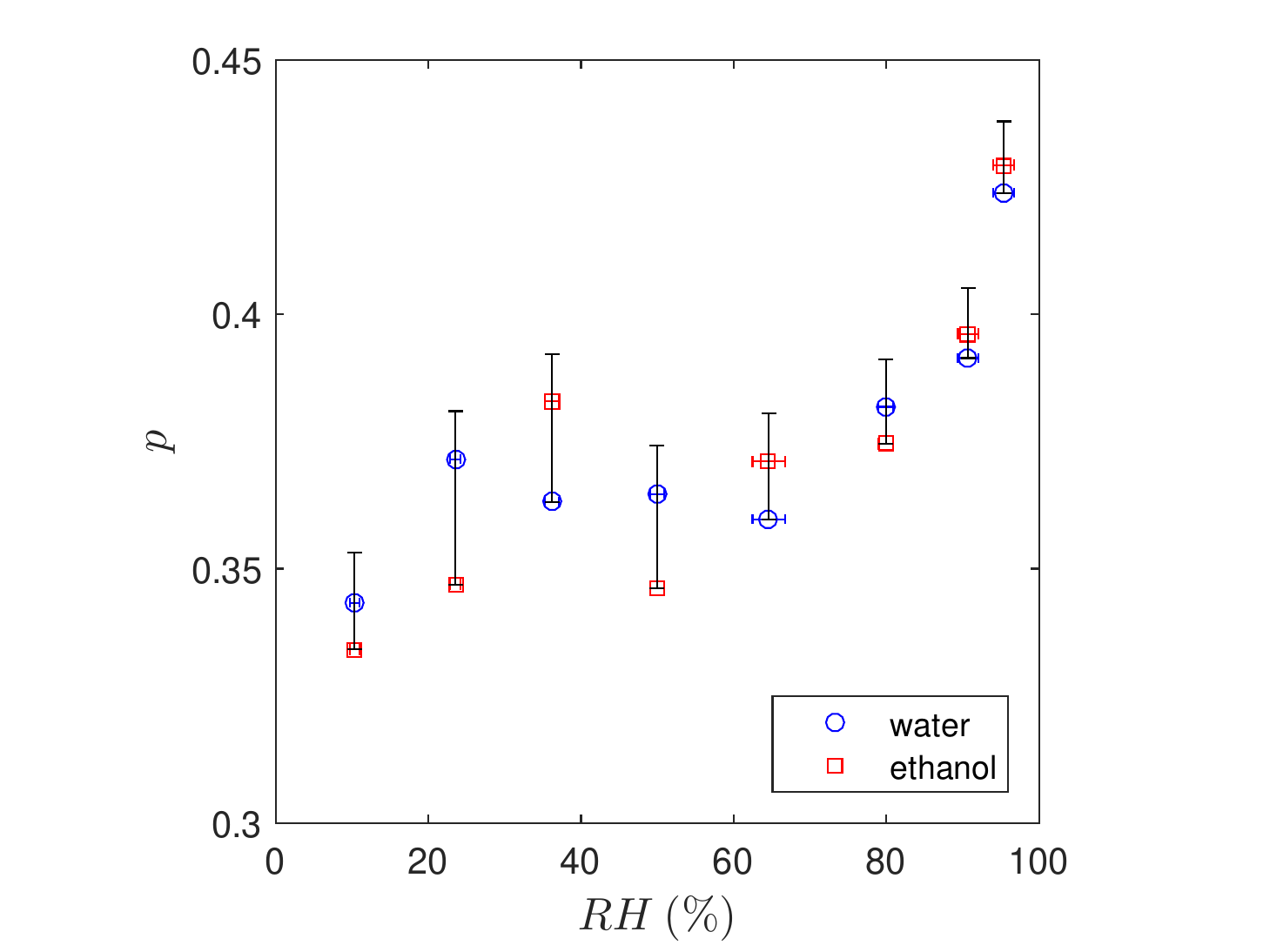}
\caption{Porosity ($p$) as a function of {relative humidity  (\RH)}, measured by hydrostatic weighing in water (blue circles) and in ethanol (red squares). 
In the rest of the article, $p$ is taken as the average of these two values. The vertical error bars account for the possible presence of water trapped in the dry samples (Tab.~\ref{tab:poro_values}).
The horizontal error bars represent the standard deviation on {\RH} during the drying experiments.
}
\label{fig:poro_vs_RH}
\end{figure}

The porosity, $p$, of the dried layer is a function of its density, $\rho$, as follows: $p= 1 - \rho/\rho_s$. Hence to determine $p$, we measure $\rho$ using classical hydrostatic weighing techniques \citep{PirLazGau16}.
In the experiments herein, the prescribed  constant relative humidity {\RH} in the enclosure is varied between $10\%$ and $95\%$, monitored within 3$\%$, {providing dried solid layers with increasing porosity, $p$, ranging between 0.34 and 0.45, respectively \citep{PirLazGau16}. 
A larger range of porosities cannot be obtained without additional expensive heat treatments \cite{Moro13PhD}. {Moreover, heat treatments may change the physical properties of the system ({\it e.g.} by inducing grain growth).}
Figure \ref{fig:poro_vs_RH} depicts the porosity as a function of \RH, and Tab.~\ref{tab:poro_values} presents the corresponding tabulation of the values.

Elastic properties (bulk, shear and Young's moduli and Poisson's ratio) are inferred via ultrasonic techniques \citep{Bar15}. Sound velocities for compression waves ($c_L$) and shear waves ($c_T$) were measured on the samples in ambient conditions, using a single transducter as described  in the appendix. Bulk modulus ($k$) and shear modulus ($\mu$) {of the macroscopic sample can then} be expressed as a function of $\rho$, $c_T$ and $c_L$: \begin{equation}
\mu = \rho c_T^2
\quad \un{and} \quad
k =\rho c_L^2 - \frac{4}{3} \mu
\label{eq:velocitytoelast}
\end{equation}

The material constants were obtained by averaging the values obtained with the transducers applied once at the top and once at the bottom of the sample. 
Also, as the resulting dried layers contain multiple fractured pieces (Fig.~\ref{fig:RH50_layer}A), the measurements were conducted on two different morsels for each drying {\RH}.

Figures \ref{fig:k_mu_vs_RH}A and \ref{fig:k_mu_vs_RH}B  present $k$ and $\mu$ as a  function of the {\RH} value for two morsels,  and Tab.~\ref{tab:elast_values} numerates the mean values. 
For humidities greater than $\sim30\%$, $k$ and $\mu$ are nearly the same in the two morsels;
this highlights the homogeneity 
{of layers dried under high humidity}. 
At low humidities ({\it i.e.}~$RH<30\%$, high evaporation rates), inhomogeneities arise, and thus, $k$ and $\mu$ differ somewhat in the two morsels.
{The observation of a drying front crossing the layer during drying for {$RH=15$ and $20 \%$} suggests these low {\RH} samples undergo directional horizontal drying. This leads to variations in the mechanical properties of these samples.}

\begin{figure}[h!]
\centering
\includegraphics[width=1\columnwidth]{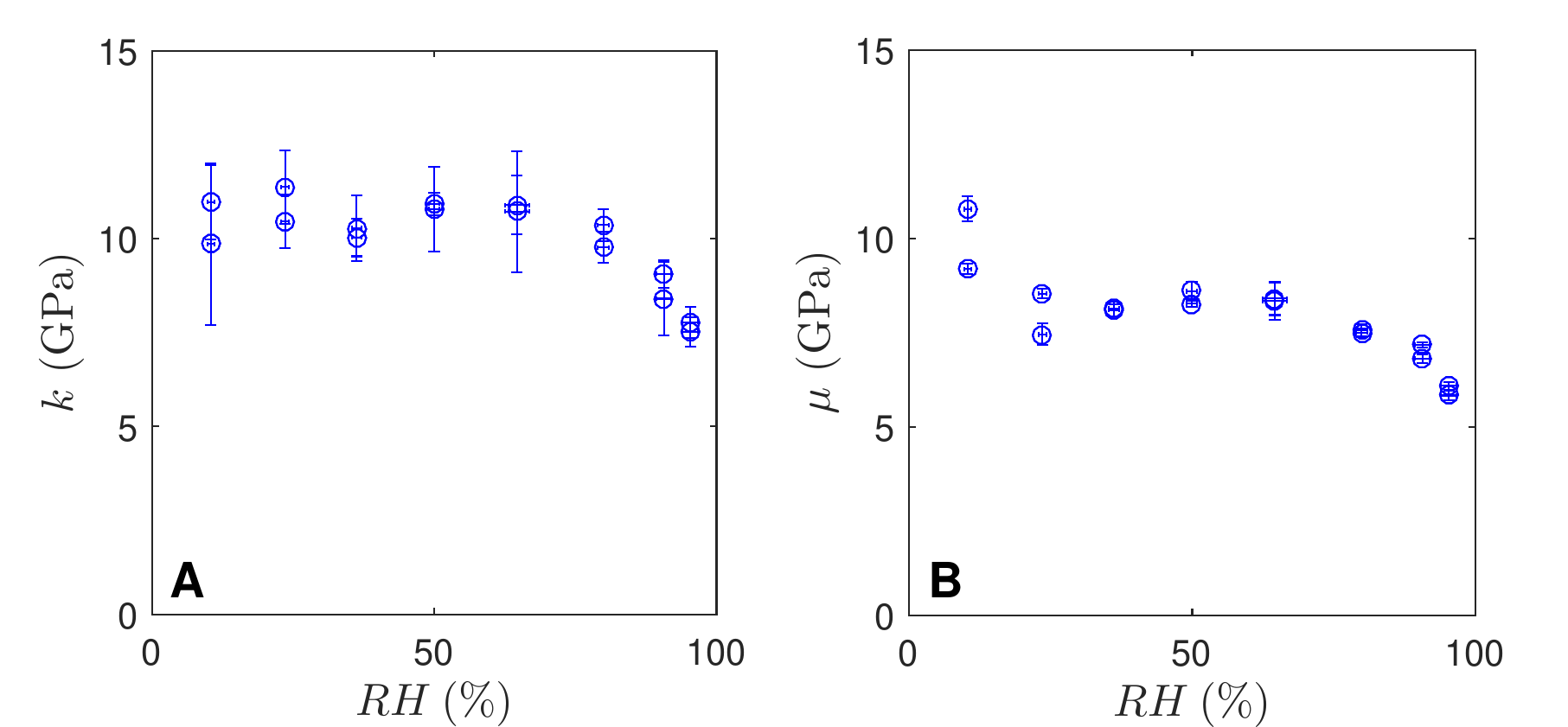}
\caption{Bulk modulus ($k$) and shear modulus ($\mu$) as a function of relative humidity (\RH). Error bars correspond to standard deviation. For each drying rate, the measurements were repeated on two morsels. 
}
\label{fig:k_mu_vs_RH}
\end{figure}

\section{Homogenization in linear elasticity}
\label{se:theory}


Since the diameter, $a$, of the colloidal particles is several decades smaller than the thickness, $h$, of the layer  ($a/h \sim 10^{-8}/10^{-3} \sim 10^{-5}$), 
one can define a Representative Elementary Volume (REV) of typical size $\ell$ such that $a \ll \ell \ll h$ ({\it i.e.}~separation of length scales hypothesis). The {scaling up} of the elastic properties from the scale of the beads to the one of the layer is thus achievable by  homogenization approaches \cite{SanZao87Wil, Pon02_Cours_Homogeneisation, DorKonUlm06}
 derived in the framework of continuum mechanics, in line
with the seminal works of Eshelby  \cite{Esh57}, Hill \cite{Hil65}, Budinsky \cite{Bud65} and Mori-Tanaka \cite{MorTan73}. 
This section summarizes the underlying hypotheses and the main results of these approaches. 

This paper approaches the problem as follows: the elastic behavior is first defined at the scale of the particles (Sec.~\ref{sse:micro}). Then, the Representative Elementary Volume (Sec.~\ref{sse:REV}) is introduced to provide a gateway between the micro- and macro- scales. 
The {scaling up} involves the resolution of the linear elasticity problem on the REV (Sec.~\ref{sse:A}). This problem is too complex to be solved  analytically, and 
approximate results are instead invoked. Section \ref{sse:mixture} summarizes the most commonly used rule of mixtures. As will be seen in Sec.~\ref{se:comp}, this simple rule does not provide an accurate estimate of the elastic properties of the porous material.

The end of the section reviews Eshelby's (Sec.~\ref{sse:eshelby}) and Mori-Tanaka's approximations (Sec.~\ref{sse:MT}) as they are prerequisites for the self-consistent scheme (Sec.~\ref{sse:sc}).
A summary of the different models is given in Sec.~\ref{se:summary}.

\subsection{Microscale}
\label{sse:micro}
{For any linear elastic material, the relation between the local strain $\tepsilon$ and stress $\tsigma$ tensors can be written as}:
$$
\tsigma(\vx)=\Cs:\tepsilon(\vx)
$$
where $\Cs$ is a proportionality {constant} expressing the elastic properties of the solid. 
Since $\tepsilon$ and $\tsigma$ are  second-order tensors, this proportionality constant takes the form of a fourth-order tensor, called the stiffness tensor.

In the case of an isotropic material,  $\Cs$ reduces to:
\begin{equation}
\C_{s}=3 k_{s} \J + 2 \mu_{s} \K
\end{equation}
with $\J$ and $\K$ respectively the spherical and deviatoric parts of the fourth-order symmetric identity tensor $\I$, given by:
\begin{equation}
 \J= \frac{1}{3} \Id \otimes \Id
 \quad \un{and} \quad
 \K= \I -\J
 \end{equation}
with $\Id$ the second-order identity tensor.
This formulation presents the advantage of being compact, and  it decouples the elastic properties into bulk $k_{s}$ and shear $\mu_{s}$ contributions. 

Notice combining the previous three equations equates to the more commonly used relation:
\begin{eqnarray*}
\tsigma=k_{s} (\mbox{tr } \tepsilon) \Id +2 \mu_{s} \te
\end{eqnarray*}
where $\te \equiv \tepsilon - \frac{1}{3} (\mbox{tr }\tepsilon) \Id $ is the deviatoric part of $\tepsilon$.
The above equations can be inverted, such that the local strain ($\tepsilon$) is a function of the stress ($\tsigma$), and is written as follows: 
\begin{eqnarray*}
\tepsilon=\frac{1+\nu_{s}}{E_{s}}\tsigma -\frac{\nu_{s}}{E_{s}} (\mbox{tr }\tsigma) \Id
\end{eqnarray*}
where the following relations between $k_{s}$, $\mu_{s}$, the Young's modulus ($E_{s}$), and the Poisson' ratio ($\nu_{s}$) are invoked:
\begin{equation}
k_{s}=\frac{E_{s}}{3(1-2 \nu_{s})}
\quad \un{and} \quad
\mu_{s} = \frac{E_{s}}{2(1+\nu_{s})}
\label{eq:kmu_Enu}
\end{equation}

At the microscale, the system can be described as a composite material made of silica (stiffness $\Cs$) and empty pores  (stiffness $\Cp$). 
In this instance, describing the elastic properties at the microscale requires a local fourth-order stiffness tensor:
$\Cloc(\vx)$  where $ \Cloc(\vx)=\Cs$ corresponds to the beads and $\Cloc(\vx)=\Cp$ corresponds to the pores, such that:
\begin{equation}
\sigma(\vx)=\Cloc(\vx) :\tepsilon (\vx), \quad \forall \vx \in \mbox{REV}
\label{eq:defCloc}
\end{equation}
Note, the stiffness of the empty pores is ideally  zero.  Yet in order to complete the calculations, it is necessary to assume a finite elasticity tensor for the pores. The final solution to the problem  then corresponds to the limits
$k_{p} /k_{s} \rightarrow 0$ and $\mu_{p}/ \mu_{s}\rightarrow 0 $, where $k_{p}$ and $\mu_{p}$ denote the bulk and shear modulus of the pores \cite{ DorKonUlm06}. Henceforth,  the results are presented in this limit.

\subsection{The Representative Elementary Volume as a gateway between the micro- and macro- scales}
\label{sse:REV}

\begin{figure}[h!]
\centering
 \includegraphics[width=0.8\columnwidth]{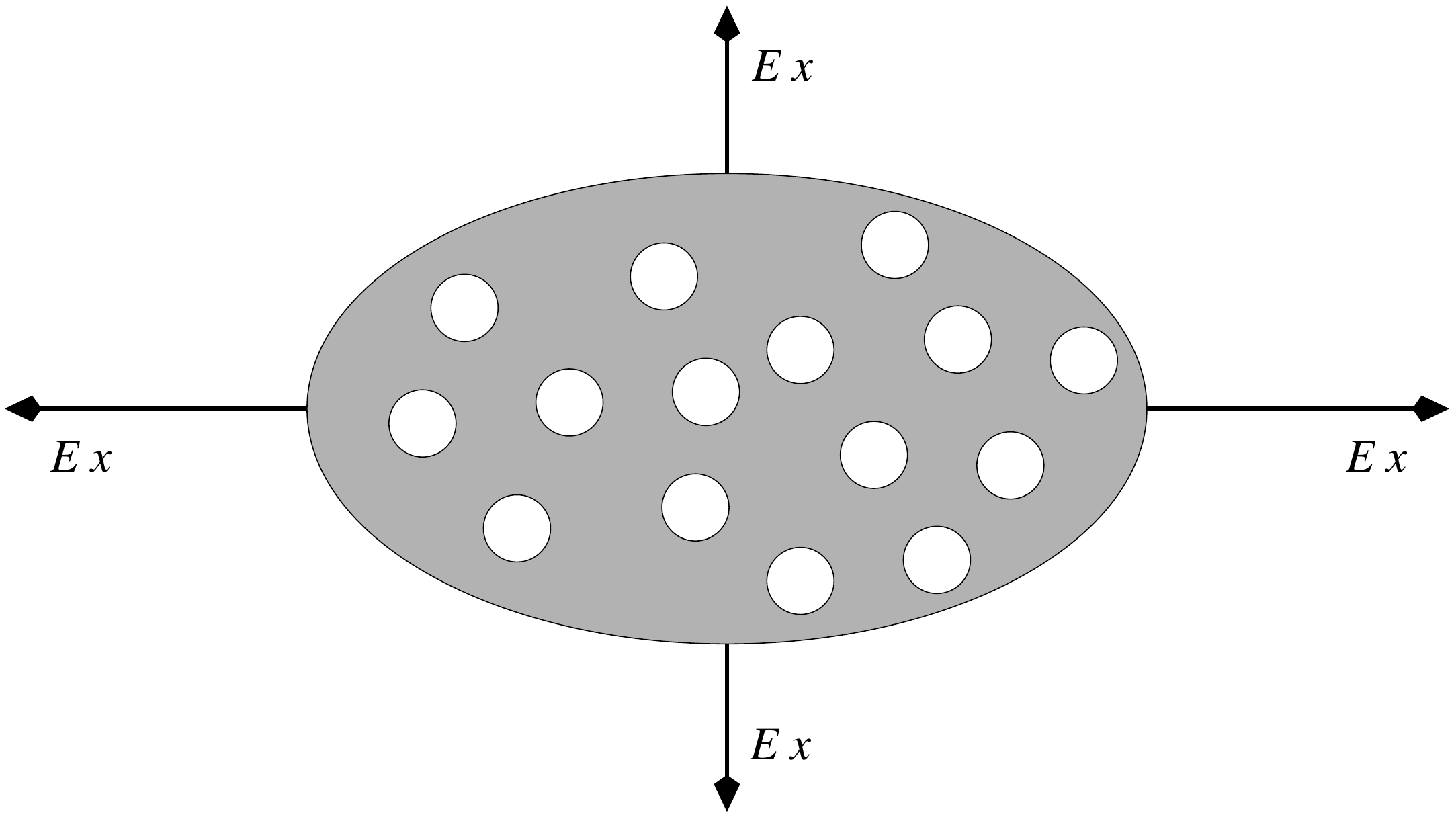}
\caption{Representative Elementary Volume (REV) corresponding to a homogeneous matrix with spherical pores submitted to displacement $\tE \vx$ along its boundary, with $\tE$ the strain tensor and $\vx$ the position vector.}
\label{fig:REV}
\end{figure}

A classical way to proceed \cite{Pon02_Cours_Homogeneisation, DorKonUlm06} is to isolate a Representative Elementary Volume (REV) and to prescribe some given displacement $\vu$  at its boundary: $\vu=\tE.\vx$, where   $\vx$ corresponds to the position vector (Fig.~\ref{fig:REV}). 
The second-order strain tensor $\tE$ corresponds to the macroscopic given strain tensor, which is assumed to be constant along the boundary. 
Using  Gauss's divergence theorem, one can demonstrate that:
\begin{equation}
\tE= \MV{\tepsilon(\vx)} 
\label{eq:defE}
\end{equation}
where $\MV{.} \equiv \frac{1}{V}\int_{REV} . \ddroit V$ denotes the spatial average over the REV.
The macroscopic stress $\tSigma$ is then defined by its mean value: 
 \begin{equation}
\tSigma\equiv\MV{\tsigma(\vx)}
\label{eq:defSigma}
\end{equation}
in coherence with the Hill-Mandel's lemma \cite{Hil67, Man72}. 

{At the macroscale, the porous material behaves as a linear elastic solid: thus, there is an effective stiffness tensor $\C$  such that 
\begin{equation}
\tSigma = \C:\tE
\label{eq:macro}
\end{equation}

If the microstructure arrangement is isotropic ({\it i.e.}~there is no preferential orientation in the way the beads are packed),   equivalent bulk ($k$) and shear ($\mu$) moduli  can be defined by:
\begin{equation}
\C=3 k \J + 2 \mu \K
\label{eq:Ckmu}
\end{equation}

The aim of homogenization is to derive the effective tensor $\C$ {knowing $\Cs$ and the microstructure arrangement.} If isotropy is valid at all scales (the material components and their arrangement are isotropic), this is equivalent to look for the relations between ($k$, $\mu$) and ($k_{s}$, $\mu_{s}$).

\subsection{The strain concentration tensor, $\A$}
\label{sse:A}
The local strain in the material  $\tepsilon(	\vx)$ represents a solution of the problem of linear elasticity defined on the REV, hence it linearly depends on the macroscopic strain ($\tE$) prescribed at the boundary of the REV. In other words, there exists a fourth-order local  tensor $\A$, called the localisation or strain concentration tensor, such that: 
\begin{equation}
\tepsilon(\vx)=\A(\vx): \tE, \quad \forall \vx \in \mbox{REV}
\label{eq:defA}
\end{equation}

{Plugging this last relation into \eqref{eq:defCloc}, taking its mean value, implementing the definition  of $\tSigma$ ({\it i.e.}~eqn \eqref{eq:defSigma}), and linking the result with the macroscopic constitutive relation (\ref{eq:macro}) leads to:}
\begin{equation}
\C=\MV{\Cloc(\vx):\A(\vx)}
\label{eq:CClocA}
\end{equation}

{Thus to obtain $\C$, it is necessary to determine $\A$, which requires solving the elasticity problem for a REV composed of a beads assembly. 
This could, {\em a priori}, be done using computationally expensive numerical schemes. Alternatively, approximate solutions may be used.  In what follows, three different estimations of $\C$ are provided: the  upper Voigt bound and the Eshelby and Mori-Tanaka estimations as prerequisites for the self-consistent approximation.

\subsection{ Voigt's upper bound \cite{Voi1889}: the rule of mixtures}
\label{sse:mixture}
A first possible approximation is to take $\A(\vx)=\I$ that is $\tepsilon(x)=\tE$ in all the REV, even in the pores.  Then,

 \begin{equation}
\C=(1-p)\Cs
\label{eq:mean}
\end{equation}

 This approximation neglects the effect of the pores on the strain distribution. It corresponds to a material behavior in which the pores deform as if they were made of the same homogeneous material as the solid phase, hence this estimation overestimates stiffness.  
One can rigorously demonstrate  that it is an upper bound for the stiffness - the Voigt bound \cite{Voi1889}.

For an isotropic material, eqn \eqref{eq:mean} is equivalent to:
\begin{equation}
\begin{array}{ccc}
 k/k_{s} &=& 1 - p\\ 
\mu/\mu_{s}&=& 1 - p 
\end{array}
\label{eq:mixture}
\end{equation}

\subsection{Pores embedded in a matrix}

Another possible approach is to consider that the packing of beads is equivalent to an homogeneous solid matrix of elasticity $\Cs$, containing a  repartition of spherical pores with the same porosity $p$ (Fig.~\ref{fig:REV}).

\subsubsection{Weak porosity: Eshelby's approximation\cite{Esh57}.}
\label{sse:eshelby}
When the porosity is weak ($p \ll 1$), the interactions between the pores can  be neglected.
Then $\A$ can  be obtained from the analytical solution of the classical problem of Eshelby: an infinite elastic matrix of stiffness  $\Cs$ containing a single spherical inclusion of stiffness $\Cp$. It is then possible to show that: 

\begin{equation}
\begin{array}{ccc}
 k/k_{s} &=& 1 - \frac{p}{1-\alpha_{s}}\\
\mu/\mu_{s}&=& 1 - \frac{p}{1-\beta_{s}}
\end{array}
\label{eq:eshelbykmu}
\end{equation}
where the expressions of $\alpha_{s}$ and $\beta_{s}$ as a function of $k_{s}$ and $\mu_{s}$ are given by:
\begin{equation}
\begin{array}{c}
\alpha_{s}=\frac{1}{1+\frac{4 \mu_{s}}{3 k_{s}}} \quad \mbox{ and }\quad
\beta_{s}=\frac{6(1+2\frac{\mu_{s}}{k_{s}})}{5(3+4\frac{\mu_{s}}{k_{s}})}  
\end{array}
\label{eq:Hill}
\end{equation}

It is noteworthy that $\alpha_{s}$ and $\beta_{s}$ depend on $k_{s}$ and $\mu_{s}$ only through  $\mu_{s}/k_{s}=\frac{3(1-2\nu_{s})}{2(1+\nu_{s})}$ (see eqn \eqref{eq:kmu_Enu}), hence only on the dimensionless Poisson's ratio $\nu_{s}$.

\subsubsection{High porosity: Mori-Tanaka's scheme\cite{MorTan73}.}
\label{sse:MT}

For interacting pores, the Eshelby approach is not sufficient: a higher-order approximation is necessary. In this case, the Mori-Tanaka scheme is relevant. 
 In this scheme, a single pore is isolated as an Eshelby problem, and a new boundary condition, $\vu=\tE_0 .\vx$,   is adopted at infinity by 
choosing an auxiliary strain tensor $\tE_0$ which takes into account the interactions between the pores. 

The detailed calculations are quite complex \cite{DorKonUlm06} and eventually lead to the following equations for $k$ and $\mu$:

\begin{equation}
\begin{array}{ccc}
 \ds\frac{k}{k_{s}}  &=&\ds \frac{1-p}{(1-p)+\frac{p}{1-\alpha_{s}}}\\
\ds\frac{\mu}{\mu_{s}} &=& \ds \frac{1-p}{(1-p)+\frac{p}{1-\beta_{s}}} 
\end{array}
\label{eq:MT}
\end{equation}
with $\alpha_{s}$ and $\beta_{s}$ given by eqn \eqref{eq:Hill}, so that the equations explicitly yield   $\frac{k}{k_{s}}$ and $\frac{\mu}{\mu_{s}}$ as a function of $p$, knowing $k_{s}/\mu_{s}$, or equivalently, $\nu_{s}$.

Notice that   Eshelby's  approximation   (\ref{eq:eshelbykmu}) corresponds to the  first order asymptotic expansion  of the Mori-Tanaka scheme (\ref{eq:MT}) with respect to  $p \ll 1$.

\subsection{Self-consistent approximation \cite{Hil65}}
\label{sse:sc}
{The Eshelby and Mori-Tanaka schemes assume a particular material geometry, in which individual, separate pores are surrounded by a solid matrix. But in the case of a sphere packing, pores are interconnected and percolate through the medium; thus, no surrounding matrix can be identified.} 
Using a self-consistent scheme, which considers an Eshelby inclusion (solid or pore) in an equivalent {homogenized} medium of stiffness $\C$,  seems preferable.
The approach used {in solving} this problem then {invokes} the one used for the Mori-Tanaka scheme, applied on the equivalent material  instead of the solid phase.  Again the derivations are  quite complex \cite{DorKonUlm06}  and only the final result is presented:

\begin{equation}
\begin{array}{ccc}
\ds\frac{k}{k_{s}} &=& \ds \frac{1-p}{1+\left(\frac{k_{s}}{k}-1\right)\alpha}\\
 \ds\frac{\mu}{\mu_{s}}&=& \ds \frac{1-p}{1+\left(\frac{\mu_{s}}{\mu}-1\right)\beta}
\end{array}
\label{eq:SC}
\end{equation}
with $\alpha=1/(1+\frac{4 \mu}{3 k})$ and $\beta=6(1+2\frac{\mu}{k})/5(3+4\frac{\mu}{k})$ as in eqn \eqref{eq:Hill}. Equation \ref{eq:SC} corresponds to two nonlinear and coupled implicit equations on $k$  and  $\mu$, which can be solved numerically, for instance using Matlab. 

\subsection{Summary}
\label{se:summary}
\begin{figure}[h!]
\centering
\includegraphics[width=0.8\columnwidth]{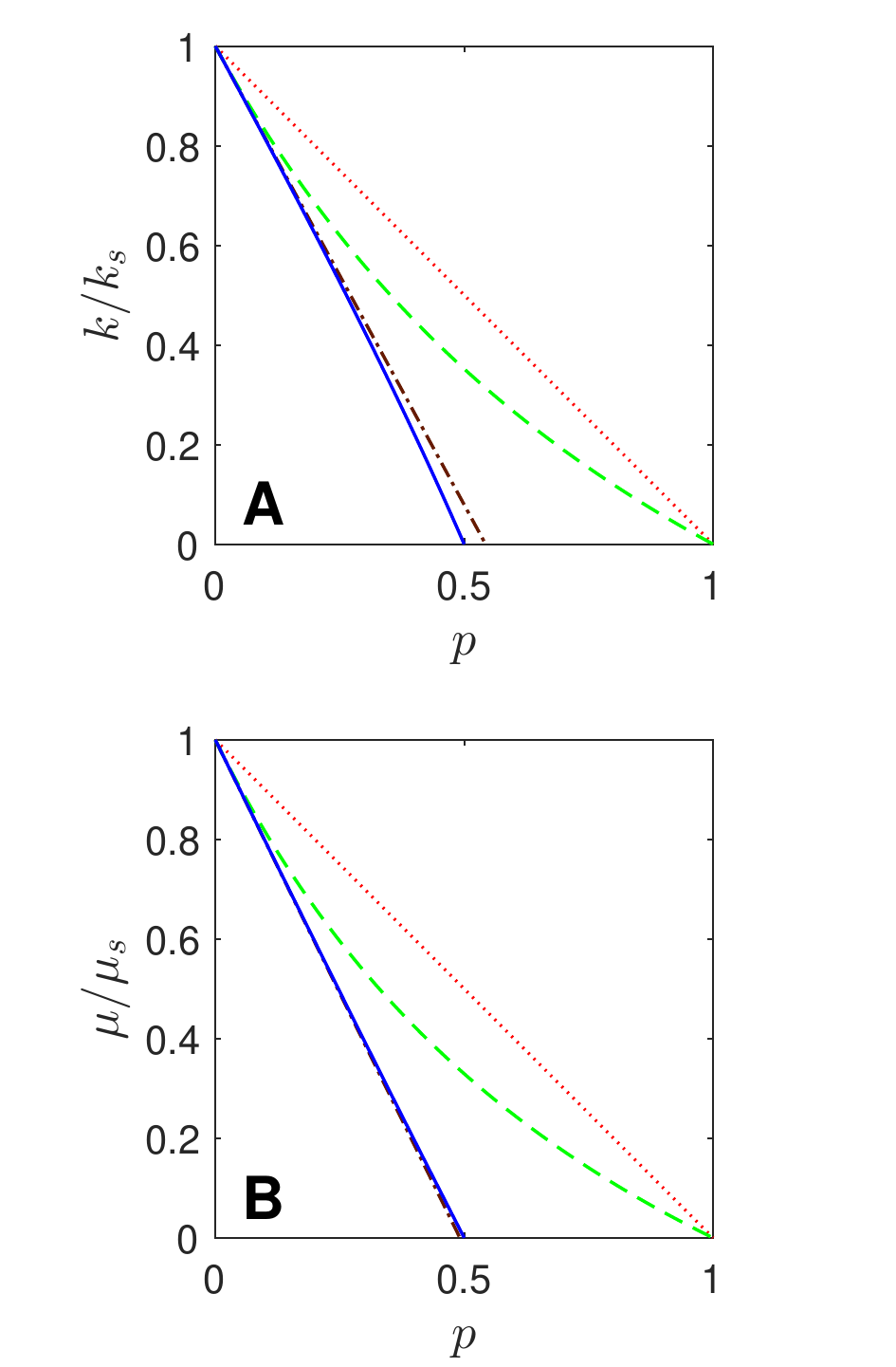}
\caption{Predictions of the different schemes for $\nu_s = 0.16$ (pure silica): rule of mixtures (red dotted line), Eshelby's approximation (brown dash-dot line), Mori-Tanaka scheme (green dashed line) and self-consistent scheme (blue solid line).}
\label{fig:three_schemes}
\end{figure}
Figure \ref{fig:three_schemes} compares the predictions of the above models for a solid phase made of silica ($\nu_s = 0.16$).
As was expected, the rule of mixtures gives an upper bound, and is thus larger than the other predictions.
All models predict a decrease in the effective stiffness with increasing porosity.
{Moreover, for a solid with negligible porosity ($p \sim 0$), all models predict moduli equivalent to the bulk parameters ({\it ie.} $k = k_s$ and $\mu = \mu_s$).}

One striking feature is that Eshelby's approximation and the self-consistent scheme yield nearly the same values. This is linked to the specific value of $\nu_s$ used here. 
In the general case, the Eshelby's approximation and the self-consistent scheme give different predictions. Moreover, the Eshelby's approximation assumes non-interacting pores and thus should not be used on materials with porosities greater than a few percent. Thus the Eshelby's model will not be further considered in this paper.}

The Mori-Tanaka scheme corresponds to} {a solid matrix containing non-connected pores. Such a system retains some stiffness} even in the vicinity of $p \rightarrow 1$, so that the Mori-Tanaka scheme predicts positive values of the moduli for $0 \leq p < 1$. 
On the contrary, the self-consistent scheme predicts that the stiffness vanishes for $p\sim 0.5$. Mathematically, it is linked to the fact that the phases invert their roles at $p=0.5$. Physically, this level of porosity is generally considered  to correspond to the percolation threshold of pores \cite{BriDorKon10}.

\section{Kendall's type models}
\label{se:kendall}

The above homogeneization schemes do not capture the effects of possible surface forces; these are expected to become significant for sub-micrometer particles, as this is the case here. It then becomes interesting to resort to Kendall's approach \citep{AlfBirKen87, Ken01}. This approach is based on the idea that the beads (of diameter $a$) are pressed together due to attractive surface forces, characterized by an interfacial energy, $W$. This energy induces a compressive force between two beads in contact \citep{JohKenRob71} :

\begin{equation}
F_{adh} = \frac{3\pi}{2}Wa
\label{eq:adh_force}
\end{equation} 

\noindent Using Hertz's theory on the contact of elastic spheres, Kendall relates the shrinkage displacement ($\delta_{adh}$) between the centers of the beads to the adhesion force ($F_{adh}$):

\begin{equation}
\delta_{adh} = \left(\frac{3}{32}\frac{1-\nu_s^2}{E_s}\frac{F_{adh}}{a^{1/2}}\right)^{2/3}
\label{eq:hertz}
\end{equation} 

\noindent Now, adding a perturbing force $f \ll F_{adh}$ to the adhesion force gives $F = F_{adh} + f$. To the first order, the perturbing force induces an additional displacement $\delta$:

\begin{equation}
f = k_n \delta \quad \mathrm{with} \quad
k_n = \left(\frac{9}{16} \frac{\pi W E_s^2 a^2}{(1-\nu_s^2)^2}\right)^{1/3}
\label{eq:pair_stiffness}
\end{equation} 

\noindent  The next step is to infer, from the stiffness of a single contact, the Young's modulus of the overall packing. Kendall first considers a simple cubic (sc) packing such that:

\begin{equation}
E_{sc} =  \frac{f/a^2}{\delta/a} = \frac{k_{n}}{a}
\label{eq:E_sc_kendall}
\end{equation} 

\noindent and the porosity of this packing is $p = 1-\pi/6 \simeq 0.4764$. Along the same lines, he then computes both the Young's modulus and the porosity for a variety of packing geometries (cubic-tetrahedral, tetragonal-sphenoidal and hexagonal) and $E$ as a function of $p$ fits well a $(1-p)^{4}$ dependance on $p$:

\begin{equation}
\begin{split}
E = A(1-\nu^2_s)^{-2/3}(1-p)^4 \left(\frac{W E_s^2}{a}\right)^{1/3} \\
\label{eq:E_kendall}
\end{split}
\end{equation} 

\noindent where $A \simeq 16.1$ is a fitting parameter \citep{AlfBirKen87}. Later, Thornton \citep{Tho93Kendall} provided some modifications of this approach: he used the theory from Johnson, Kendall and Roberts \citep{JohKenRob71} (rather than Hertz theory) to estimate contact stiffness, and he considered a body centered orthorombic array (rather than simple cubic) as a reference packing. These two modifications reduce $A$ by a factor of two. In a nutshell, $A$ varies from 8 to 16.1, {depending on the model and its assumptions.}

\section{Experimental results versus theoretical predictions}
\label{se:comp}
\subsection{Comparison with homogenization approaches}
{Figures \ref{fig:k_mu_poro}A and \ref{fig:k_mu_poro}B respectively report $k/k_{s}$ and $\mu/\mu_{s}$ as a function of $p$, where the datapoints corresponds to those of Fig.~\ref{fig:poro_vs_RH}. These figures also present the predictions of the different models discussed in Sec.~\ref{se:theory}.}
{As anticipated in Sec.~\ref{se:theory}, the rule of mixtures (eqn \eqref{eq:mixture}) significantly overestimates the experimental data whereas the self-consistent model (eqn \eqref{eq:SC}) accurately predicts them. 
The Mori-Tanaka scheme (eqn \eqref{eq:MT}) is observed to}  fall in between.

\begin{figure}[h!]
\centering
\includegraphics[width=0.8\columnwidth]{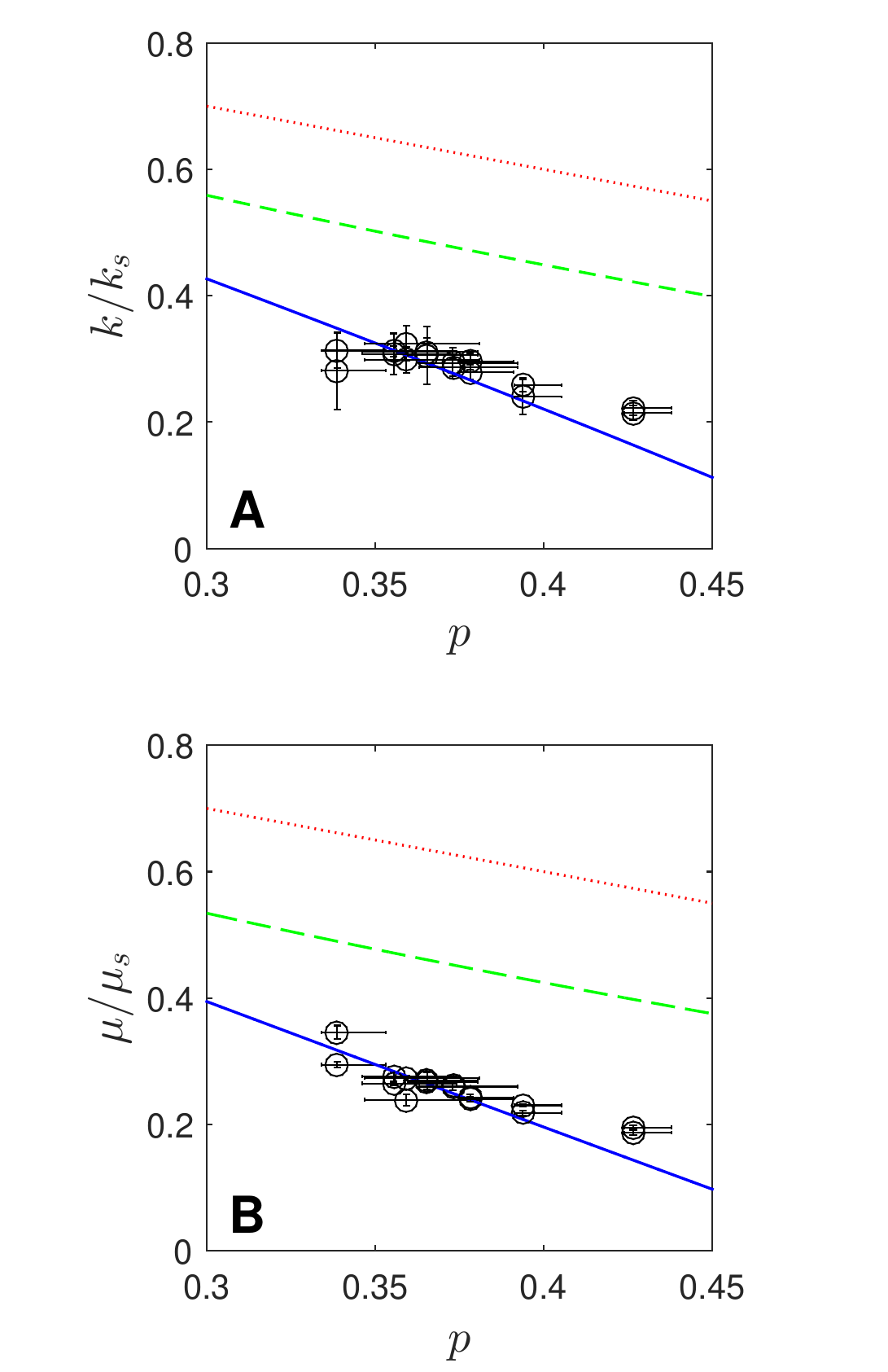}
\caption{Normalized bulk modulus ($k/k_s$) and shear modulus ($\mu/\mu_s$) as a function of porosity ($p$). Normalization procedure uses bulk fused silica glass parameters (spec values from Tab. \ref{Atab:silica}). Lines correspond to theoretical schemes: rule of mixtures (red dotted line), Mori-Tanaka scheme (green dashed line) and self-consistent scheme (blue solid line).}
\label{fig:k_mu_poro}
\end{figure}

{The agreement between theory and experiment} for both bulk and shear constants {supports} the hypothesis of {isotropy} made in the paper.
{If  significant anisotropy existed in the packing, it would have resulted in the invalidation of eqn \eqref{eq:Ckmu} and subsequent results.}
{\it De facto}, this gives some indication  of the {3D packing structure} which is difficult to access otherwise.

{
For the sake of completeness, 
Figures \ref{fig:E_nu_poro}A and \ref{fig:E_nu_poro}B present the Young's modulus ($E$) and Poisson's ratio ($\nu$) as a function of $p$ using the following equations \citep{LanLif70}:}
\begin{equation}
E = \frac{9 k \mu}{3 k + \mu}
\quad \un{and} \quad
\nu = \frac{3k-2\mu}{2(3k+\mu)}
\label{eq:velocitytoelast2}
\end{equation} 

\begin{figure}[H]
\centering
\includegraphics[width=\columnwidth]{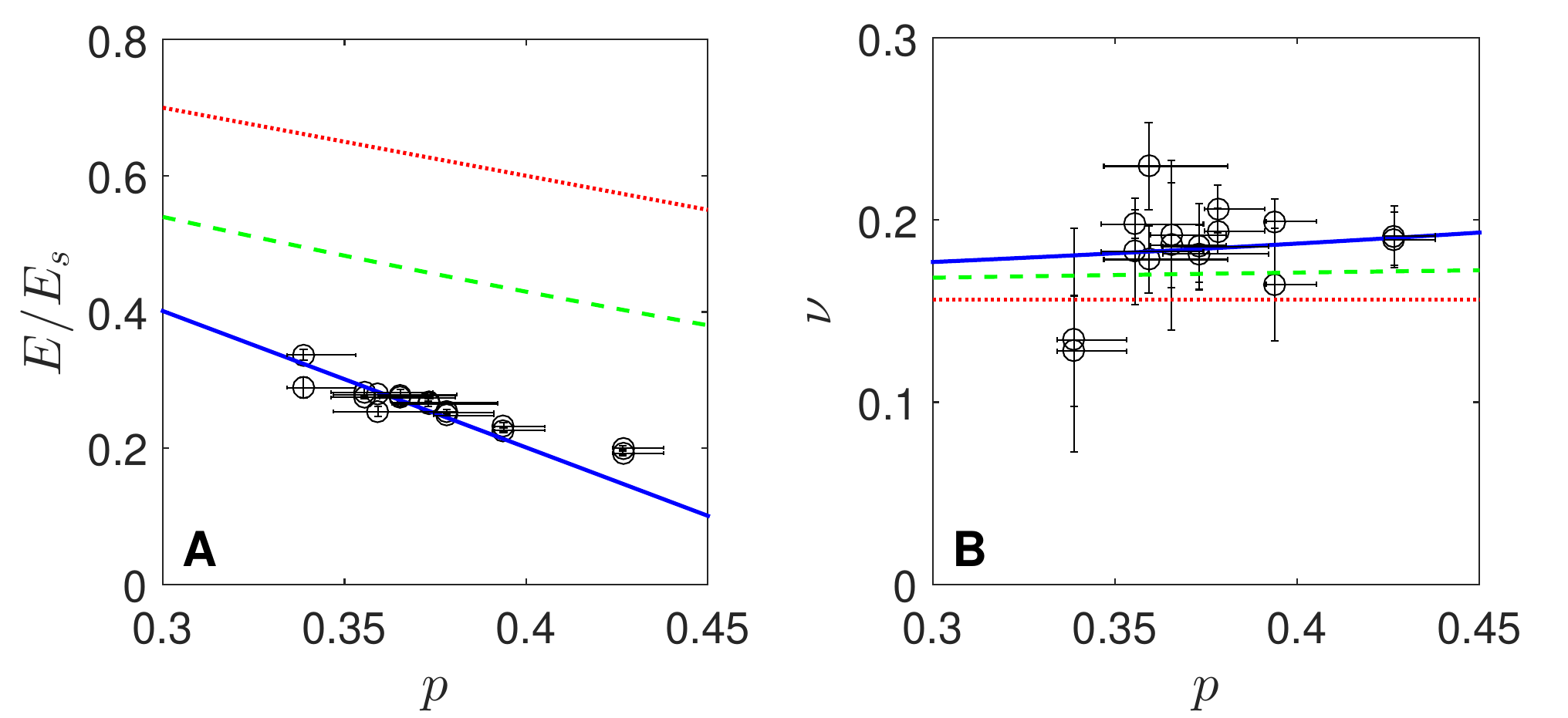}
\caption{Normalized Young modulus ($E$) and Poisson's ratio ($\nu$) as a function of porosity ($p$). The experimental values  for $E$ were normalized by the spec value of bulk fused silica glass ($E_s$). Lines correspond to theoretical schemes: rule of mixtures (red dotted line), Mori-Tanaka scheme (green dashed line) and self-consistent scheme (blue solid line).}
\label{fig:E_nu_poro}
\end{figure}

Porosity is observed to affect only $E$, while $\nu$ remains constant within the errorbars. The latter can be interpreted in conjunction with recent works \cite{Rou07,GreaGreLak11} correlating $\nu$ with the short-to-medium range connectivity of the network at the microscale, which, hence, likely remains almost constant for the considered porosity range.

\subsection{Comparison with Kendall's type approach}
\begin{figure}[h!]
\centering
\includegraphics[width=0.8\columnwidth]{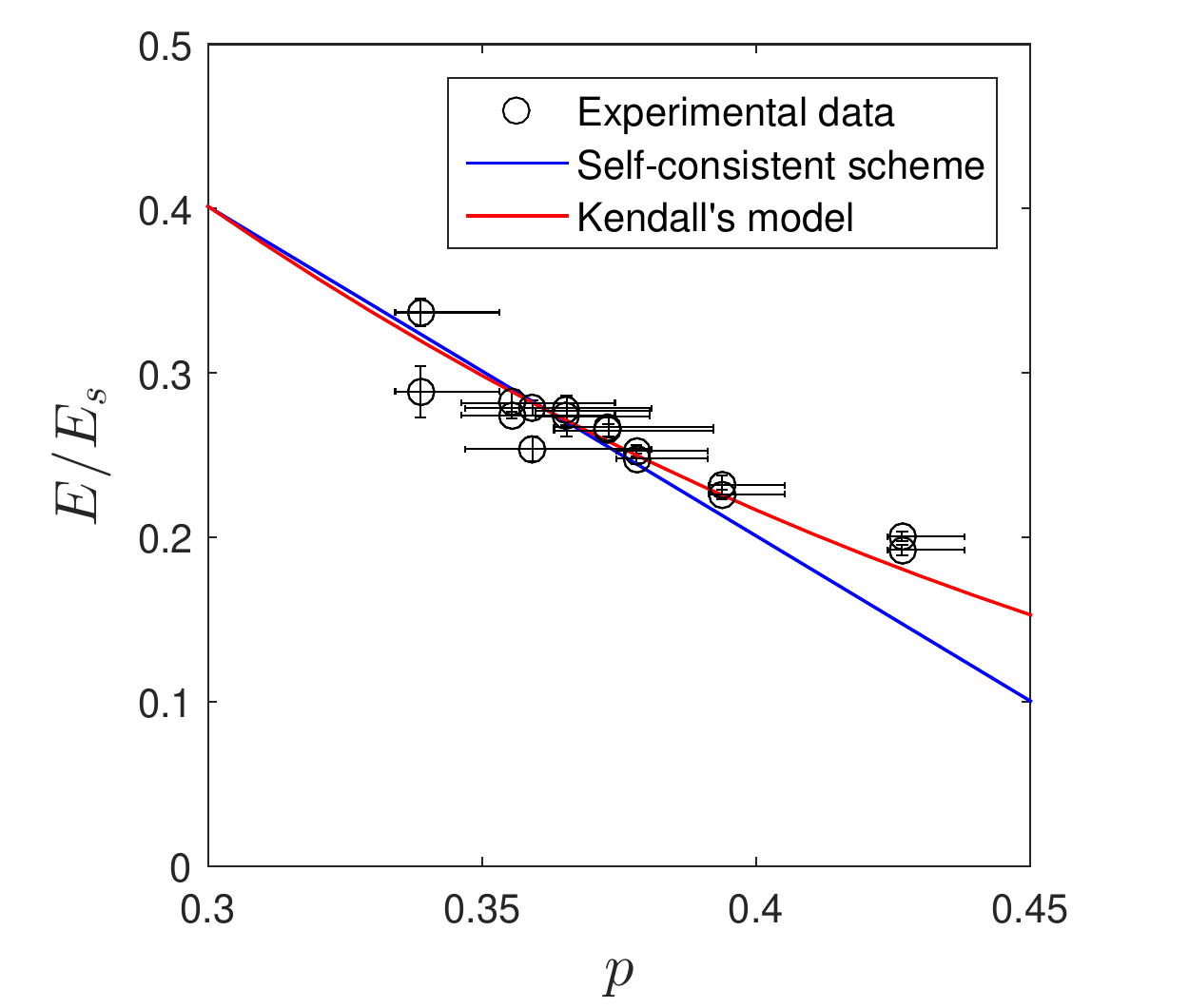}
\caption{Normalized Young's modulus as a function of porosity. The numeric values used in Kendall's expression (eqn \eqref{eq:E_kendall}) were $E_s = 72.7 \un{GPa}$, $a = 16.2 \un{nm}$ (measured values) and $W = 5 \un{J/m}^2$ (fitted value for $A = 10.3$, as used in \citep{CleGoeRou13, BirYunRee17,GoeNakDut15DessicationCracks}).}
\label{fig:E_kendall}
\end{figure}
Figure \ref{fig:E_kendall} compares the experimentally measured Young's modulus with the prediction of the self-consistent scheme (eqns. \eqref{eq:SC} and \eqref{eq:velocitytoelast2}) and that of Kendall (eqn \eqref{eq:E_kendall}). The latter better describes the dependency on $p$ for $p>0.35$, at the price of an additional fitting parameter: the adhesion energy $W$.  Taking, as before, $E_s=72.7 \un{GPa}$ and $a=16.2 \un{nm}$, the fit of the experimental data with Kendall's equation gives $W = 1.1 \un{J/m}^2$ and $W = 10.7 \un{J/m}^2$ for $A = 16.1$ and $A = 8$ respectively.
Now,  two types of surface energies are potentially relevant in the problem, as the interfacial energy could be attributed either to the breaking of covalent bonds (here siloxane) or to a surface tension (here silica/air or silica/water) as in fluids \cite{CleGoeRou13, BirYunRee17}.  The  macroscopic fracture energy for silica $G_{c} \simeq 8.2 \un{J/m}^{2}$  \cite{RouBon14} 
{relates to the breaking of silica bonds, as does Griffith's energy \cite{Gri20} $2 \gamma_s \sim 3.8 \un{J/m}^{2}$. This energy represents the work applied to break the bonds on a unit surface: this is inferred} 
from the energy needed to break a single bond\cite{Luo07}, $7.5\,10^{-19} \un{J/bond}$ ($450 \un{kJ/mol}$), assuming 5 bonds per nm$^{2}$ \cite{Ile79} and no subsequent dissipation mechanisms such as plasticity or damage. The energies {relating to the surface tension} are considerably weaker: $\gamma_{SV} \sim 0.3-0.5 \un{J/m}^2$ for silica/vacuum interfaces \cite{Sar06} and about six times less for silica/water interfaces \cite{CleGoeRou13}.

The fitted value for $W$ is several orders of magnitude larger than the ones associated with surface tension, but in line with the fracture energy of silica. This demonstrates that sintering during drying has {probably} led to the formation of covalent siloxane bonds between the particles.

\section{Discussion}
\label{se:discussion}
\subsection{Competing models}

The main outcomes from Sec.~\ref{se:comp} are recalled and discussed:
\begin{itemize}
\item The self-consistent scheme provides the complete (tensorial) description of the elasticity, with no adjustable parameter. This for instance provides information on the isotropy of the material.
\item The self-consistent scheme predicts an almost linear dependence of elastic constants with porosity. This is in agreement with the experiments for $p<0.35$. At higher porosities, a small discrepancy with the experimental data appears, which suggests a missing physical ingredient. 
\item Kendall's model succeeds in reproducing the behavior of the Young's modulus over the whole porosity range studied herein. Thus, the discrepancy between the self-consistent scheme and the experimental data is likely due to surface effects, which are accounted for in Kendall's model.
\item The increased accuracy of Kendall's model comes at the cost of an additional adjustable parameter, the adhesion energy. Moreover, the value of the prefactor involved in the model ($A$ in eqn \eqref{eq:E_kendall}) {depends on the precise assumptions made in the model} \cite{Tho93Kendall} and has a tremendous effect ($A^3$ dependence) on the determination of the adhesion energy. 
\item Kendall's model only predicts the Young's modulus and cannot provide any information on the isotropy of the material.
\end{itemize}
\noindent It should be stressed that looking for elasticity around ponctual contacts would result at the macroscopic scale in a non-linear stress-strain relationship (hertzian theory). As a consequence eqn \eqref{eq:defA} (and thus eqn \eqref{eq:macro}) and eqn \eqref{eq:E_kendall} remain valid provided that only small perturbations around an unloaded reference configuration with a {\it finite} contact area are considered.
 Finally, as the two approaches are not mutually exclusive, it may be possible to refine the self-consistent scheme   
{to take into account the adhesive surface forces \cite{BriDorKon10,BriDorKon10bis} as in Kendall's approach.}

\subsection{The physics of colloidal drying}

This study involves particles with 15 nm diameter. It is expected that surface effects decrease with increasing particle size, hence the self-consistent scheme is expected to be more accurate (over a larger porosity range) as the particle size increases (and vice versa). Additional experiments with particles of different diameters should verify this hypothesis.

The occurrence of cracks during drying reveals the cohesive nature of the packing. These cracks are due to the combination of three effects: the adhesion between the particles and the substrate, the overall retraction induced by water evaporation, and the transmission of tensile stresses from the substrate throughout the whole layer via the adhesion between the particles.
Interpretating the present data in light of Kendall's model provides an estimate of the adhesion energy.
\begin{itemize}
\item The value inferred here for the adhesion energy ($W \simeq 1 - 10 \un{J/m}^{2}$) is on the order of the fracture energy commonly reported for silica \cite{Rou07}. Hence covalent siloxane bonds were {probably} formed between the particles during the drying. 
\item  This value is orders of magnitude larger than the one used in \cite{CleGoeRou13,BirYunRee17}  ($W=\gamma \simeq 0.01 \un{J/m}^{2}$) to describe the same system (dried Ludox HS-40). However their study concerns the onset of cracking, while our measurements {concern dried samples}. This suggests an {significant} evolution of $W$ with time during the drying process.
\item  The value of the adhesion energy can be recast into a fracture toughness using Irwin's relation \cite{Irw57}: $K_c = \sqrt{E W} \simeq 0.15 - 0.5 \, \un{MPa.m}^{1/2}$. This value measured at the end of drying is in line with the one estimated in \cite{GauLazPau10} at the onset of cracking. 
\end{itemize}

 A challenge is to measure {\it in situ} and independently, from the onset of cracking to the end of drying, the evolution of both the elastic properties and the fracture energy. This would provide important information on how and when the covalent bonds are formed.

\section{Conclusion}

To summarize, this paper uses a highly porous material formed by the drying of a colloidal suspension as a benchmark for homogeneization schemes of mechanical behavior. 
Using ultrasound measurements, it investigates the elastic properties of a dried layer of silica nanospheres. 
By modulating the drying rate, the influence of the porosity has been studied. 
It has been
demonstrated that 
the self-consistent scheme  accurately predicts  both elasticity constants, with no ajustable parameters, as long as  the porosity {is} small enough (less than $35\%$ for $15 \un{nm}$ beads).
For higher porosities, surface effects become visible; Kendall approach succeeds in taking them into account for a single elastic constant (Young's modulus), at the price of an additional ajustable parameter (adhesion energy). 
This adhesion energy is found to be on the order of the fracture energy in silica, implying that at the end of the drying the beads are probably linked by covalent bonds.

A surface energy equal to the fracture energy of the pure silica leads to a fracture toughness in line with that reported in \cite{GauLazPau10} for this kind of system. 
{A step forward in these works would consist of measuring the elastic properties during drying.  This would provide information on the formation of covalent bonds.}
Further steps are  (i)  to test self-consistent and Kendall's predictions by varying the particle size as this is expected to change the contribution of the surface effects and
(ii) to use Kendall's approach 
to interpret, in physical terms, recent mathematical refinements of self-consistent schemes  \cite{BriDorKon10, BriDorKon10bis}.

\section{Acknowledgments}

This work has benefited from the financial support of the  Institut Universitaire de France,  Triangle de la Physique (RTRA), Ile-de-France (C'Nano and ISC-PIF),  LabEx LaSIPS (ANR-10-LABX-0040-LaSIPS) and PALM (ANR-10-LABX-0039-PALM) managed by the French National Research Agency under the "Investissements d'avenir" program (ANR-11-IDEX-0003-02).
We thank D.~Kondo and L.~Dormieux for encouragements and helpful discussions about the homogenization schemes. 
We   thank M.~Imp\'{e}ror for the SAXS characterisation of the initial suspension.

\begin{table}[H]
  \centering
  \caption{Measured values for the porosity in each sample, using hydrostatic weighting in water and in ethanol. Due to water retention in the pores, actual values may be up to $1.5\%$ larger \citep{PirLazGau16}.}
    \begin{tabular}{ccc}
    \hline
    $RH$    & $p_{w}$ & $p_{eth}$ \\
    \hline
    $11$  & $0.343$ & $0.334$ \\
    $23$  & $0.372$ & $0.347$ \\
    $36$  & $0.363$ & $0.383$ \\
    $50$  & $0.365$ & $0.346$ \\
    $65$  & $0.36$ & $0.371$ \\
    $80$  & $0.382$ & $0.375$ \\
    $90$  & $0.391$ & $0.396$ \\
    $95$  & $0.424$ & $0.429$ \\
    \hline
    \end{tabular}%
  \label{tab:poro_values}%
\end{table}%

\begin{table}[H]
  \centering
  \caption{Measured values for the elastic constants in each sample. Measurements for each sample were repeated on two morsels.}
    \begin{tabular}{ccccc}
     \hline
    $RH$    & $k$ (GPa) & $\mu$ (GPa) & $E$ (GPa) & $\nu$ \\
     \hline
    $11$  & $10.97 \pm 0.99$ & $10.78 \pm 0.34$ & $24.31 \pm 0.59$ & $0.13 \pm 0.04$ \\
    $11$  & $9.85 \pm 2.15$ & $9.19 \pm 0.14$ & $20.84 \pm 1.13$ & $0.13 \pm 0.07$ \\
     \hline
    $23$  & $11.37 \pm 0.98$ & $7.45 \pm 0.29$ & $18.32 \pm 0.54$ & $0.23 \pm 0.03$ \\
    $23$  & $10.46 \pm 0.72$ & $8.53 \pm 0.13$ & $20.11 \pm 0.34$ & $0.18 \pm 0.02$ \\
     \hline
    $36$  & $10.28 \pm 0.87$ & $8.13 \pm 0.2$ & $19.27 \pm 0.43$ & $0.19 \pm 0.03$ \\
    $36$  & $10.03 \pm 0.51$ & $8.1 \pm 0.17$ & $19.13 \pm 0.29$ & $0.18 \pm 0.02$ \\
     \hline
    $50$  & $10.79 \pm 1.14$ & $8.6 \pm 0.24$ & $20.34 \pm 0.54$ & $0.18 \pm 0.03$ \\
    $50$  & $10.92 \pm 0.3$ & $8.26 \pm 0.08$ & $19.8 \pm 0.16$ & $0.2 \pm 0.01$ \\
     \hline
    $65$  & $10.72 \pm 1.61$ & $8.34 \pm 0.5$ & $19.75 \pm 0.89$ & $0.19 \pm 0.05$ \\
    $65$  & $10.89 \pm 0.79$ & $8.4 \pm 0.45$ & $20.01 \pm 0.65$ & $0.19 \pm 0.03$ \\
     \hline
    $80$  & $10.36 \pm 0.44$ & $7.56 \pm 0.16$ & $18.23 \pm 0.27$ & $0.21 \pm 0.02$ \\
    $80$  & $9.76 \pm 0.42$ & $7.5 \pm 0.12$ & $17.9 \pm 0.23$ & $0.19 \pm 0.02$ \\
     \hline
    $90$  & $8.4 \pm 0.99$ & $7.18 \pm 0.08$ & $16.72 \pm 0.45$ & $0.16 \pm 0.04$ \\
    $90$  & $9.07 \pm 0.37$ & $6.81 \pm 0.13$ & $16.33 \pm 0.22$ & $0.2 \pm 0.02$ \\
     \hline
    $95$  & $7.78 \pm 0.42$ & $6.09 \pm 0.11$ & $14.47 \pm 0.22$ & $0.19 \pm 0.02$ \\
    $95$  & $7.52 \pm 0.4$ & $5.83 \pm 0.14$ & $13.89 \pm 0.23$ & $0.19 \pm 0.02$ \\
     \hline
    \end{tabular}%
  \label{tab:elast_values}%
\end{table}%




\providecommand*{\mcitethebibliography}{\thebibliography}
\csname @ifundefined\endcsname{endmcitethebibliography}
{\let\endmcitethebibliography\endthebibliography}{}

\bibliographystyle{apsrev4-1} 

\appendix

\section{Sound velocities measurements}
\label{A:ultra}

This appendix details methods for acquiring the elastic moduli and Poisson's ratio of the dried colloidal layers via ultrasonic techniques. As the morsels are thin ($\sim 2\un{mm}$), this requires special signal processing techniques in calculating longitudinal ($c_L$) and transverse ($c_T$) wave speeds.

A single transducer is coupled to a face of the sample, using honey as the viscous couplant for pulse transmission. The transducer coupled with a pulse generator (Panametrics 5800)  either provides a compression (Olympus Panametrics-NDT M116)  or shear (Olympus Panametrics-NDT V222) pulse to the sample, the frequency of the pulse is $20\un{MHz}$ in both cases. 
The ultrasonic  system provides a controlled short pulse so as to control the  wave introducted into the sample:  $\sim 0.4~\mathrm{\mu}\text{s}$ for both compressional and shear pulses. 

The emitted pulse travels across the specimen, bouncing back and forth between the two opposite faces. Its successive passings at the specimen/transducer interface are detected at $500\un{MHz}$ using a Tektronic TDS3054B oscilloscope. Figure \ref{Afig:raw_signal} display the tension captured by the oscilloscope as a function of time. The thickness of the specimen is measured using a digital caliper with accuracy $0.01\un{mm}$.

\begin{figure}[H]
\centering
\includegraphics[width=0.8\columnwidth]{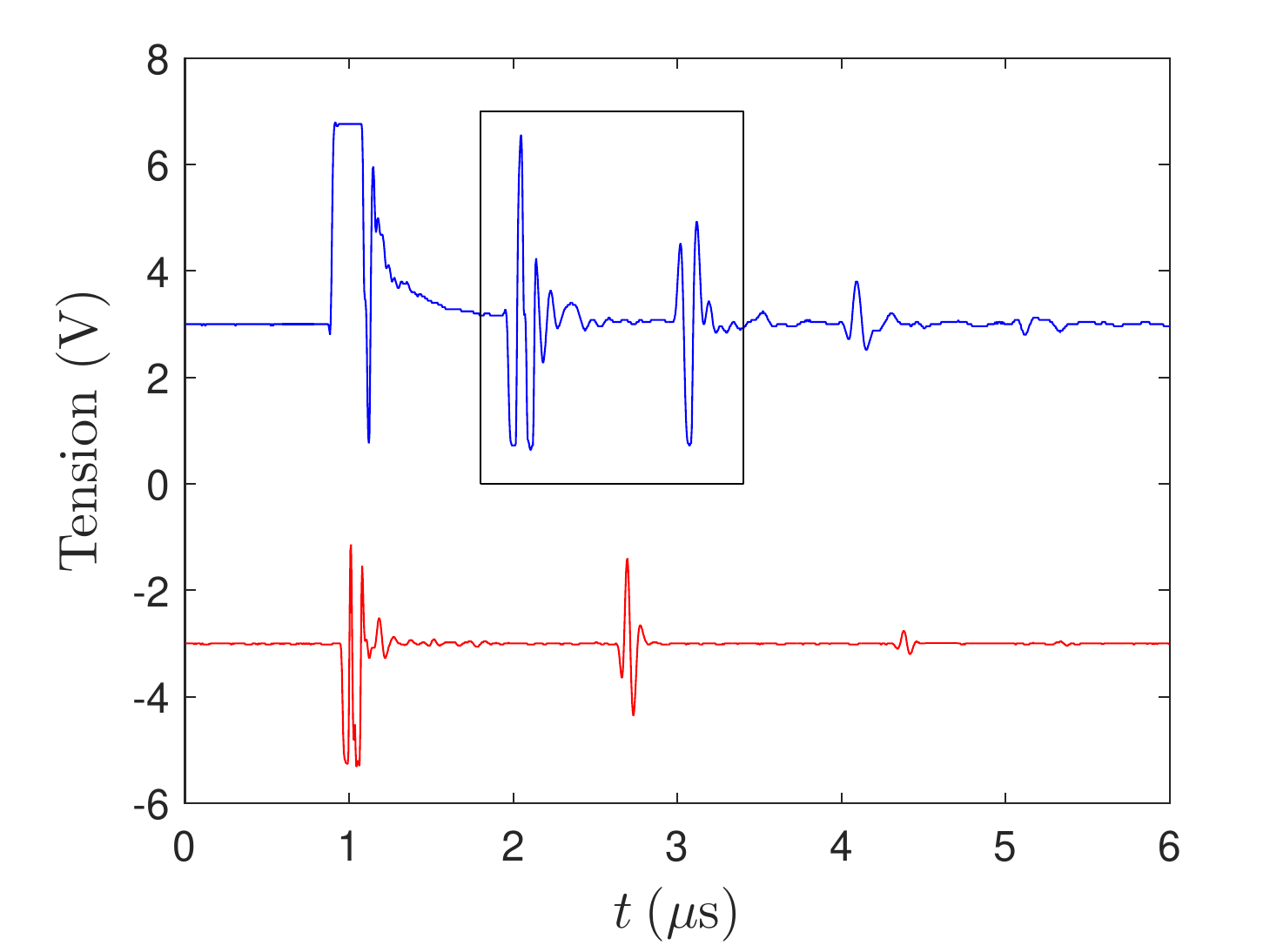}
\caption{Time evolution of the signal received at the transducer for compression waves (blue) and shear waves (bottom). The material probed here is a sample dried at $RH=50\%$ and the specimen thickness is $2.07 \pm 0.03 \un{mm}$. For the sake of clarity, the signals are shifted by $3\un{V}$ and $-3\un{V}$ respectively. The black box delimits the signal represented on Fig.~\ref{Afig:matching_peaks}.}
\label{Afig:raw_signal}
\end{figure} 
 
In our experiment,  the samples are thin ($\sim 2\un{mm}$): the typical time between two successive echoes is $\sim 1~\mathrm{\mu}\text{s}$ (resp. $\sim 1.6~\mathrm{\mu}\text{s}$) for compressional (resp. shear) pulses. In general, associating a single discrete time to the pulse arrival on the transducer proved difficult. This difficulty was overcome by the following processing scheme:
\begin{itemize}
\item For each pulse, the arrival times of three successive extrema of the signal are determined.
\item For each series of matching extrema, the time delay between two matching extrema (ie extrema of the same color on Fig.~\ref{Afig:matching_peaks}) gives an estimate of the propagation time of the pulse through the sample. 
\item These estimates are evaluated between each pulse for each series of matching extrema, and subsequently averaged (Fig.~\ref{Afig:t_avg}).
\end{itemize}
\noindent Bulk modulus ($k$) and shear modulus ($\mu$) can then be deduced  from $c_L$, $c_T$ 
and the density ($\rho$) 
using eqn \eqref{eq:velocitytoelast}. \\

\begin{figure}[H]
\begin{center}
\includegraphics[width=0.8\columnwidth]{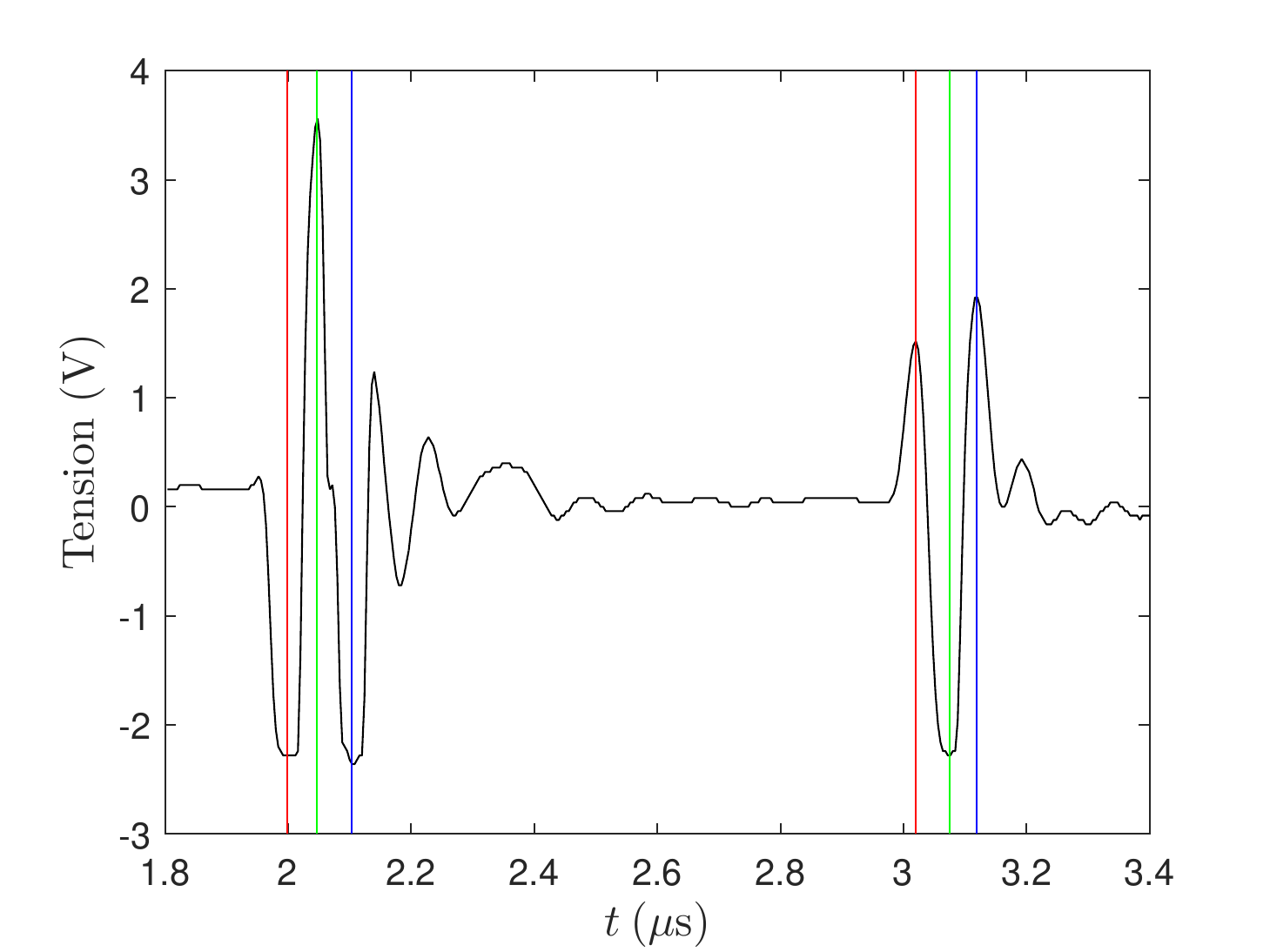}
\caption{Two successive reflected pulses for the compression wave represented on Fig.~\ref{Afig:raw_signal}. Arrival times of matching extrema are indicated by vertical lines of identical color.}
\label{Afig:matching_peaks}
\end{center}
\end{figure}
   
\begin{figure}[h]
\begin{center}
\includegraphics[width=0.8\columnwidth]{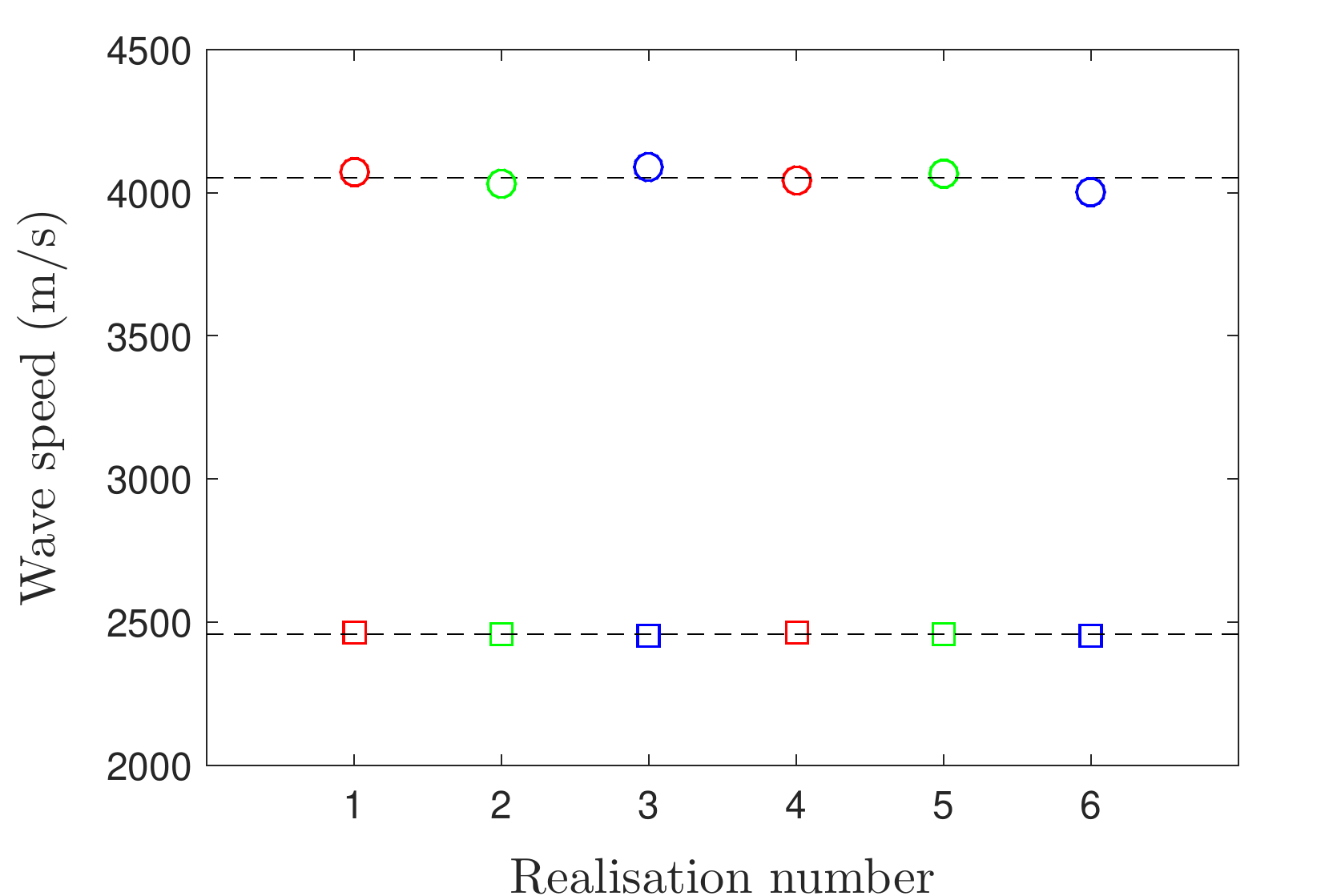}
\caption{Time delay over travelled distance for each pair of matching extrema, between pulses 1 and 2 and pulses 2 and 3, for compression waves (circles) and shear waves (squares). The black dotted lines represents the average of all estimated speeds for a given type of waves.}
\label{Afig:t_avg}
\end{center}
\end{figure}

In order to validate this measurement procedure, we applied it to a block (5x5x25 mm$^{3}$) of pure silica glass (Corning 7980 standard grade).  
Measurements of the sound velocities were repeated on the two sets of opposite faces. 
As for the colloidal layers, density was determined by hydrostatic weighting. We measured $\rho_{s} = 2.20 \pm 0.02 \un{g/cm}^3$, in accordance with the value provided by the supplier.
The corresponding values of $k_{s}$ and $\mu_{s}$
are presented in Tab.~\ref{Atab:silica} together with the values provided by the supplier. They are in good agreement with each other.  
This validates our characterisation methods.\\

\begin{table}[H]
  \centering
    \begin{tabular}{ccc}
    \hline
          & ${k_s}$ (GPa)& ${\mu_s}$ (GPa)    \\
          \hline
    Spec. & 35.4 & 31.4  \\
    \hline
    Meas. 1 & 35.9$\pm$0.3 & 31.0$\pm$0.1  \\
    Meas. 2 & 35.8$\pm$0.8 & 31.4$\pm$0.2 \\
         \hline
    \end{tabular}%

    \caption{Elastic constants measured by ultrasounds on fused silica standard grade, Corning code 7980 ($\pm$ provide the errorbars for one standard deviation).}
        \label{Atab:silica}
    \end{table}

\end{document}